# Integrated modelling approaches for sustainable agri-economic growth and environmental improvement: Examples from Canada, Greece, and Ireland

Garcia J.A.[1], Alamanos A.[2*]

[1]The Water Institute, University of Waterloo, 200 University Avenue W, Waterloo, ON N2L 3G1, Canada. Email: ja4garci@uwaterloo.ca

[2]Department of Civil Engineering, University of Thessaly, Volos, Greece; alamanos@civ.uth.gr.

* Corresponding author: alamanos@civ.uth.gr

**ABSTRACT**

Complex agricultural problems concern many countries, as the economic motives are increasingly higher, and at the same time the consequences from the irrational resources' use and emissions are becoming more evident. In this work we study three of the most common agricultural problems and model them through optimization techniques, showing ways to assess conflicting objectives together as a system and provide overall optimum solutions. The studied problems refer to: i) a water-scarce area with overexploited surface and groundwater resources due to over-pumping for irrigation (Central Greece), ii) a water-abundant area with issues of water quality deterioration caused by agriculture (Southern Ontario, Canada), iii) and a case of intensified agriculture based on animal farming that causes issues of water, soil quality degradation, and increased greenhouse gases emissions (Central Ireland). Linear, non-linear, and Goal Programming optimization techniques have been developed and applied for each case to maximize farmers' welfare, make a less intensive use of environmental resources, and control the emission of pollutants. The proposed approaches and their solutions are novel applications for each case-study, compared to the existing literature and practice. Furthermore, they provide useful insights for most countries facing similar problems, they are easily applicable, and developed and solved in publicly available tools such as Python.

## 1. Introduction

Complex agricultural problems with conflicting economic-production and environmental objectives are becoming increasingly topical (Lu Zhang et al., 2018). Agriculture consumes environmental resources (soil, water, raw materials), emits pollution (fertilizers, pesticides, greenhouse gases – GHGs) and has high production expectations (increase yields, production, sales and profits, reduce costs). The optimal way to cover the economic demand, and achieve environmental sustainability through the most efficient use of resources and emissions' control is a challenging multi-objective problem that requires integrated approaches to balance those, often conflicting, objectives (Arias et al., 2019).

A commonly studied problem is the water quality deterioration from agricultural pollutants. Nitrogen (N) and phosphorus (P) from agricultural runoff are a major concern for many countries (Xia et al., 2020). Intensive livestock production can also be a significant source of nutrients' discharge, water bodies' pollution, and greenhouse gases emissions (Kato et al., 2009). Monitoring water quality parameters is fundamental and can lead to advanced modelling techniques, highly informative for the policy-makers, e.g. as in Pipelzadeh and Mastouri (2021), who assessed contaminants concentration with classification-based models. Such approaches are also essential to identify the main pressures within agriculture (e.g. fertilizers, pesticides, soil erosion, livestock pollution, etc.) and set priorities for their control (Chen et al., 2018). Management models using scenario analysis and/or Best Management Practices (BMPs) are widely used, as well as Decision-Support-Systems (DSS) to evaluate the different alternatives (Nicholson et al., 2020). The methods to control or treat non-point pollution refer to source control, process control, or end treatment, where the literature is vast; however, there are fewer efforts to minimize pollution by integrating the farmers' perspective (e.g. ensuring same or higher profits to the farmers) (Xia et al., 2020; Jianchang et al., 2015).

Water quantity management and allocation has also received great attention in the context of conflict resolution of different users and environmental sustainability, and the study of their trade-offs (Kilic and Anac, 2010; Knox et al., 2018). Studying and optimizing the distribution of water volume over time and space has produced several advanced modelling routines (Parasyris et al., 2021; Fu et al., 2018), and provided useful solutions for optimum allocation in water-scarce areas that rely on agriculture (Spiliotis et al., 2014; Farrokhzadeh et al., 2020). The incorporation of more factors, such as the water–energy–land–food nexus (Psomas et al., 2021), and the economy through hydro-economic models (Hashmi et al., 2019; Alamanos, 2021), has contributed to more integrated decision-making (Alcoforado de Moraes et al., 2021).

Analyzing water quality, quantity, economic, and social aspects together is more challenging in terms of conceptualization, data, modelling, and reaching in agreements (Loucks and van Beek, 2017). Hatamkhani and Moridi (2021) analyzed agricultural production, economic efficiency and social equality; Sahoo et al. (2006) studied the trade-offs and optimal planning of land-water-crop systems. Some approaches combine water quantity and quality management modelling or evaluating BMP scenarios to protect or improve water systems (Liu et al., 2021; Alamanos et al., 2019). Such combinative studies further highlight the importance of using integrated approaches to understand better the systems' interactions, uncertainties, and balance multiple competitive objectives in a more efficient way (Li et al., 2020).

This work aims to build on that direction and encourage similar efforts, despite the complex and data-hungry nature of integrated models, that have been the main limitations and impediments to their practical application so far (Delli Priscoli et al., 2004; Loucks and van Beek, 2017). We aim to show that common systemic issues can be successfully addressed in different environments, through analyzing water quality and quantity management through classic optimization techniques, combined with socio-economic aspects, supporting thus the policy-making with valuable information covering multiple sectors. Based on experiences from different case studies, three representative problems are presented, covering a wide range of issues that concern many countries: a) a dry Mediterranean watershed in Central Greece with overexploited surface and groundwater resources due to the intensified agriculture; b) a typical northern watershed in Ontario, Canada, facing water quality issues from agricultural runoff; c) and an Irish watershed, where the expanding livestock farming deteriorates water quality. The role of the proposed Linear, Non-linear, and Goal Programming optimization models in the above cases, is to balance conflicting objectives, enhance multidisciplinary approaches and knowledge, providing thus insightful points for consideration by policy-makers, and encouraging similar systemic approaches.

## 2. Study areas

### 2.1 Lake Karla Watershed (LKW), Greece

Agricultural production is a major driver of the economy of the Mediterranean regions, as their climate can support the production of various high-quality agricultural products. At the same time, their relatively dry climate stresses water availability (Ravazzoli et al., 2021; Peres and Cancelliere, 2016). Increased production, efficiency and profit objectives often lead to water resources overexploitation and mismanagement, although the literature has highlighted the necessity of a policy shift towards more sustainable and efficient solutions, considering the environmental constraints (Iglesias et al., 2011; Suárez-Almiñana et al., 2020).

In Greece, the lack of rational irrigation water management is becoming more evident as water needs are hardly covered, especially in Thessaly, the driest Basin District of the country (Loukas et al., 2007). The problem of the optimum water resources allocation and use-efficiency in the region has been approached through simulation and management models (Manakos et al., 2011; Charizopoulos et al., 2018), while a few studies applied optimization techniques for water management in case studies at Thessaly. Manos et al. (2013) used mathematical programming based on multicriteria techniques to optimize the agricultural production considering profits, agronomic, and environmental factors. Sidiropoulos et al. (2013) optimized the pumping of the overexploited aquifer of LKW. Panagopoulos et al. (2014) optimized the irrigation practices and their implementation based on their cost-effectiveness to avoid the future desertification of Pinios River Basin. Sakellariou-Makrantonaki et al. (2016) applied linear programming for the optimum management of the surface irrigation network of Pinios. Alamanos et al. (2017) presented several optimization problems for LKW using agriculture, economics and water management objectives to study their trade-offs.

LKW in Thessaly is an intensively cultivated area of 117,300 ha, with overexploited water resources (its aquifer, river Pinios, and Lake Karla) (Fig.1a). Lake Karla was drained in 1962 for flood protection and more farmland, causing several environmental problems (depletion of the aquifer, pollution of surface and groundwater resources, changes in the local climate, extreme events, etc.), leading to the restoration of the former lake from 1981 until 2011 (Sidiropoulos et al., 2013). Until today the new lake does not fully operate because the accompanying irrigation works are incomplete, there are issues of water theft (supposed to supply the reservoir from Pinios River), and the general poor management of the area. Losses from evaporation, leakage, inefficient irrigation practices, intensification of agriculture, illegal wells, lack of economic management, project planning and political will to address the conflicting agricultural and environmental sides, complete the frame with the major concerns. The farmers often protest against potential environmental measures, while subsidies and product prices have been driving the policy-making so far. So, the *optimization example presented for LKW maximizes net-profits under environmental constraints*, in an attempt to explore the degree up to which the first affects negatively the latter, for the first time in the study area.

## 2.2 Northern Lake Erie Basin (NLEB), Ontario, Canada

Ontario' Action Plans consider several strategies to improve water quality, focusing on the reduction of P runoff to Lake Erie to tackle eutrophication (Watson et al., 2016). Agriculture and agri-food manufacturing are the main pressure, but also responsible for the largest GDP in Ontario (Statistics Canada, 2020). Governmental goals for controlling nutrients' concentrations have been set, and the reduction of P runoff up to 40% is the major one (Ontario Ministry of Environment, 2010).

The main focus has been on the pollution control, rather than studying the conflicts between environmental (water quality) and economic (agricultural production) objectives (Bereket and de Loë, 2020; Alamanos and Garcia, 2021). Pollution control strategies and BMPs have been extensively studied and modelled, mainly by simulating and evaluating specific measures and scenarios (Stonehouse and Bohl, 1993; Zebarth et al., 2009; Agro and Zheng, 2014; Yang et al., 2014; Schreiner, 2017; Hanief and Laursen, 2019). The interlinkages of crop production, surface runoff and P emissions indicate the need to study them in integrated models (Wang et al., 2018). Recently, the source control of P has been noted (Liu et al., 2021), so the detailed consideration of land use management is needed (Archonditsis et al., 2019). The effect of different crop distributions to nutrient runoff loads has been increasingly highlighted; however, agricultural economy is the competing force (Michalak et al., 2013; Adhami et al., 2019). Subsequently, the research question arising is how to balance the economic-environmental conflicts – how to achieve same or higher profits with reduced nutrients' exports. The literature highlights the need to combine all affecting factors into a single model, comprehensive and useful for policymakers. On this basis, this work provides an *optimization framework to address the trade-offs of agro-economics subject to environmental (pollution) constraints*, at the NLEB, which includes 274 watersheds and sub-watersheds (2,270,000 ha) directly running off to Lake Erie (Fig.1b). Optimization has been used so far for other purposes in rural Canadian studies, such as maximizing cropping yields and production or minimizing production costs (Jeffrey et al., 1992; Liu et al., 2013). To our knowledge, this is the first application of farmers' profit maximization through the optimum crop distribution under pollution control constraints.

## 2.3 Erne Sub-Catchment Area (ESCA), Ireland

Ireland is a primarily rural country, where animal farming is a significant economic driver. Agriculture is the main cause of diffuse pollution (fertilizers, pesticides with low nutrient-use efficiency, manure and Carbon (C) emissions) (Sharpley et al., 2013). Agriculture accounts for approximately 30% of Ireland's GHGs emissions (Chiodi et al., 2015), the negative impacts of cattle access to water bodies on their quality have been extensively highlighted (Conroy et al., 2016), and all the above fundamentally disrupt P and N cycles and deteriorate water quality (Styles et al., 2006; Nasr et al., 2007).

The rural layout of the country includes many small local communities and only a few urban centres. The water management is carried out by privately (locally) voluntary owned and operated schemes, or the Group Water Schemes (GWS), which support the construction, operation and maintenance of water supply and distribution systems from local sources such as lakes or boreholes into homes and farms (Irish Water, 2021). The National Federation of Group Water Schemes (NFGWS, 2021) is the representative organisation for the community-owned group water scheme sector in Ireland. Although the EPA and the Local Authorities provide general guidelines and programmes on pollution control (better manage fertilizers, manure, etc.), the monitoring and management of pressures and measures' impact to inform relevant decisions in practice is still in a poor and primitive stage.

ESCA (Fig.1c) covers 9,840 ha and was chosen as a typical case facing agriculture's diffuse pollution with excess nutrient losses and GHG emissions, like most Irish catchments (EPA, 2018; 2021). The main information source for most areas is the River Basin Management Plans (RBMPs), a requirement of the European Water Framework Directive, where every Member State must outline measures to restore and protect the status (quantity and quality) of the water bodies, operating over 6-year cycles. A basic monitoring – characterization process is in place, using high-moderate-poor-at risk status categories, as per the Water Framework Directive. All ESCA's surface water bodies are historically of poor quality (Hayward, 1993), were characterised as 'poor status' (2010-2015), being further deteriorated since the 2007-09's assessment (EPA, 2018). The recent assessment for the 3$^{rd}$ Cycle of the RBMPs showed no improvement in the sub-catchment's water bodies (EPA, 2021) – where all the surface water bodies were characterised "at risk" (meaning that it is unlikely to achieve good ecological status by 2027). Two lakes (Kill and Graddum) and the groundwater are the drinking water sources for the catchment. The two lakes are "not on a published monitoring programme", their status is "unassigned" since 2007, while the groundwater is "under review"; and both river bodies are at risk, without any chemical monitoring (Catchments.ie, 2021). The responsible Local Authority (LA) is Cavan County Council and the main measure suggested by the RBMP refers to its planned work 'and potential to build on findings'.

Agricultural practices, the food and livestock industry development, and the pollution from agriculture is being studied and modelled by Teagasc (Agriculture and Food Development Authority), and certain Irish research centres. Very few integrated approaches have been proposed so far, focusing in a smaller and very detailed scale, e.g. Breen et al. (2019) for dairy farms. There has been limited focus on finding ways to address the conflicting issues of the agriculture's expansion and the environmental degradation at a larger (catchment scale). This work presents a *Goal Programming model considering various environmental, economic, agronomic, and social objectives* to explore their trade-offs, for the first time for ESCA and Irish agriculture.

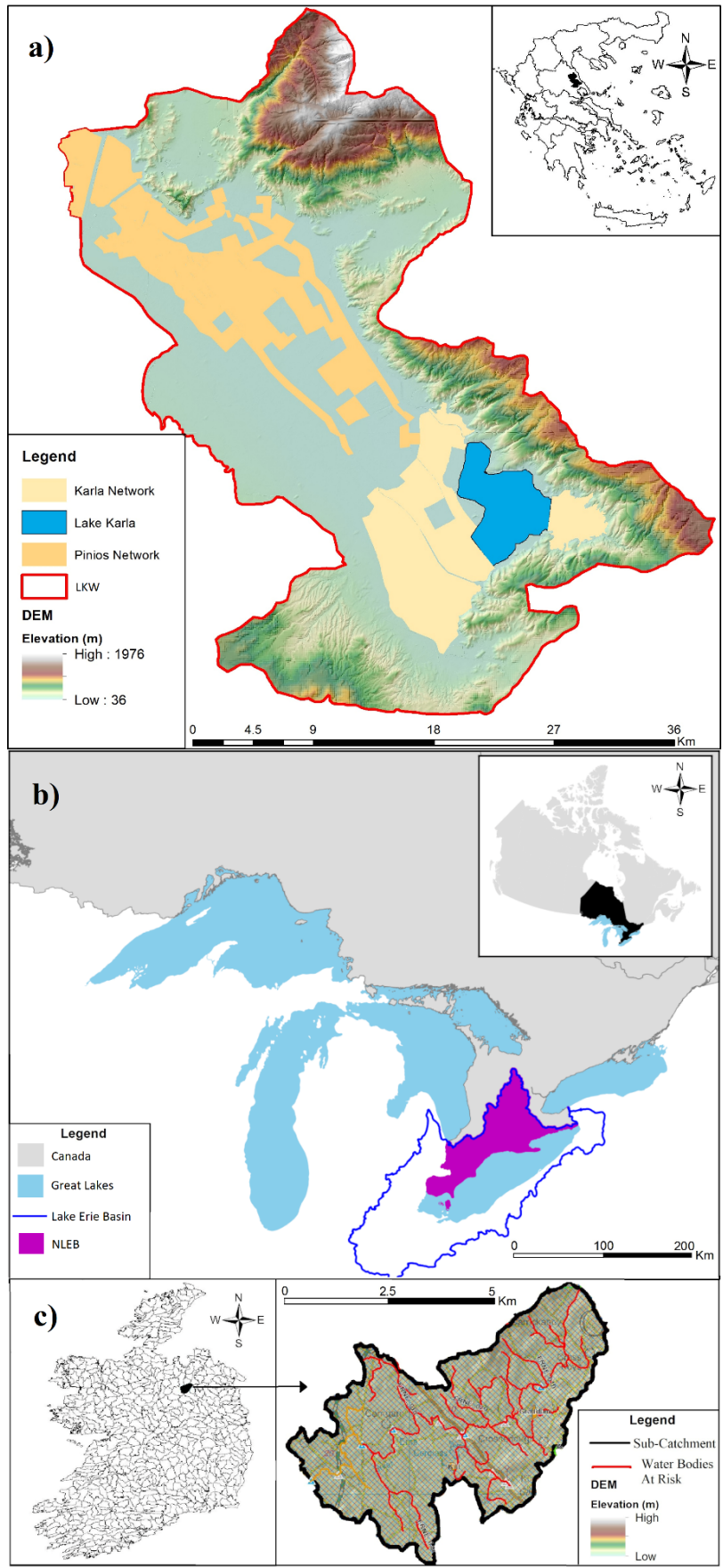

Figure 1. The three sites used for the modelling examples: a) LKW, b) NLEB, c) ESCA.

## 3. Methodology

### 3.1 Linear, Non-Linear and Goal Programming

An optimization process using linear programming assumes a linear objective function (Z) set as a goal for maximization (or minimization), under linear constraints, all functions of the decision variables (Eq.1):

$$Z_{\max \text{ (or min)}} = f(x_1, x_2, x_3, \ldots, x_n) \tag{1}$$

where $(x_1, x_2, x_3, \ldots, x_n)$ are the decision variables, the system's data. In addition, it must satisfy a set of constraints, the acceptable range of values (Eq.2):

$$u_i(x_1, x_2, x_3, \ldots, x_n) \leq a_i \tag{2}$$

where $a_i$: known values. The optimum solution of the system must meet all the constraints and the objective function.

This practically provides a useful set-up for problems like the ones described in the previous section, because an objective (goal) can be maximized or minimized, while exploiting the optimum levels of the other parameters of the system (controlled as constraints), all depending on the decision variables.

In the case of non-linear programming, the relations of Eq.(1) or (2) are described by non-linear functions. These problems are in general harder to solve; however, optimality can be guaranteed when certain conditions are met by the problem.

Goal Programming is a powerful and flexible technique that can be applied to a variety of decision problems involving multiple objectives (Charnes and Cooper, 1961). It attempts to minimize the set of deviation from multiple pre-specified (desirable) goals which are considered simultaneously. The analyst or stakeholders can weight those goals according to their importance, in a way e.g. to penalize the deviations from them (so that lower order goals are considered only after the higher order goals). The general Goal Programing model is based on linear programing model (Eq.3):

$$\text{Min } Z = \sum_{i=1}^{m} \sum_{k=1}^{K} P_k \left( w_{i,k}^+ d_i^+ + w_{i,k}^- d_i^- \right) \tag{3}$$

The goals are expressed by the 'm' component and $P_k$ is the priority coefficient for the k-th priority. '$w_i$' represents the weights of each goal and $d_i^+$, $d_i^-$ are the deviational variables representing the amount of under-achievement and over-achievement of the i-th goal respectively.

The constraints (Eq.4) are expressed through the decision variable '$x_j$' and the $b_i$ and $a_{ij}$ coefficients for the j-th decision variable in the i-th constraint.

$$\begin{aligned}
&\sum_{j=1}^{n} a_{ij} x_j + d_i^- - d_i^+ = b_i \quad for\ i = 1, 2, \ldots, m \\
&\sum_{j=1}^{n} a_{ij} x_j \ (\leq\ =\ \geq)\ b_i \quad for\ i = m+1, \ldots, m+p \\
&x_j, d_i^-, d_i^+ \geq 0 \quad for\ j=1,2,\ldots n,\ i=1,2,\ldots,m
\end{aligned} \tag{4}$$

Where $w_{i,k}^+$ the weight for the $d_i^+$ variable in the k-th priority level and $w_{i,k}^-$ the weight for the $d_i^-$ variable in the k-th priority level.

These approaches were chosen due to their relatively simple structure, suitability to the studied problems' characteristics, their expression through relations that can be easily modified if needed, and their ability to provide straightforward and clear-cut solutions in a reasonable computational time and capacity. The software used for the development and solution of the problems was Python (Anaconda), which is free and reliable for linear programming, non-linear, integer and mixed optimization problems (Kruk, 2018).

### 3.2 The LKW model

After an exploration of different models (Alamanos et al., 2017), a linear optimization problem was deemed most appropriate for the LKW's case. The model maximizes the profits from the agricultural activities, subject to the watershed's area, water, fertilizer, and labour constraints (Table 1).

Table 1. Mathematical description of the LKW model (annual values).

| Relation | Description |
|---|---|
| $x_i, \quad i = 1,\ldots,18.$ | Decision variable $x_i$ represents the area allocated to each crop $i$ in [ha]. |
| $\max NP = \sum_i np_i \cdot x_i \quad (5)$ | The objective function is the maximization of net profits in [€]. The coefficient $np_i$ represents the net profit per area of each crop [€/ha]. |
| $\sum_i x_i \leq \text{TotalArea} \quad (6)$ | 1st constraint: not to surpass the total available cultivated area [ha]. |
| $\sum_i wr_i \cdot x_i \leq \text{TWA} \quad (7)$ | 2nd constraint: the water requirements for each crop ($wr_i$ in [m³/ha]) not to exceed the total water availability (TWA), i.e., the renewable water resources [m³]. |
| $\sum_i fert_i \cdot x_i \leq \text{TF} \quad (8)$ | 3rd constraint: not to surpass the current applied fertilization quantity from all crops' requirement in [kg]. |
| $\sum_i lb_i \cdot x_i \leq \text{TL} \quad (9)$ | 4th constraint: not to exceed the total available labour hours for work on the cultivation of crop in [hr]. |
| $x_i \geq x_i^{\min} \quad (10)$ | 5th constraint sets a minimal bound of cultivated area for each crop. |

Equation (10) ensures the optimal solution do not abruptly diverge from the cultivation pattern observed during the last ten years in the area. This is ensured by setting a lower bound of cultivation area for each crop.

The NP of Table 1 was estimated using a straightforward relation of the gross margin minus the total production cost. The gross margin is the sum of total revenue (production multiplied with product prices) plus the subsidies, while the production cost (for one unit of product) is as the sum of the costs of lubrication, herbicides, seed, two sprays, defoliants, harvesting cost, pumping costs, oil, labor, planting cost, mechanical operations, and agricultural deductions, (Alamanos et al., 2019). 11 irrigated and seven dry crops were used as decision variables (xi) and their typical economic data, water, fertilizer and labour requirements were retrieved from official data sources (Appendix – Table A1). While the TF and TL were estimated using the existing crop distribution, the TWA were estimated based on the concept of the renewable water resources, in order to avoid negative water balances (water deficits), and thus environmental degradation. The water availability was estimated using the hydrological model UTHBAL (Loukas et al. 2007) to calculate, among other variables, the surface runoff and the groundwater recharge, namely, the renewable surface water and groundwater resources (Alamanos et al., 2019). This constraint is essential for this water-scarce area, as it allows the sustainable use of water resources, i.e. the renewable volume, rather than overexploiting the aquifer's stocks.

The scenarios tested on this model explored the different crop allocations that can be obtained from reducing the current water use in the basin. This was performed by reducing the baseline total water use by 2% up to 80% (covering thus a continuous range of all the possible scenarios between those values). Therefore, the model provides a crop mix that is less water intensive but still economically appealing to farmers by maximizing their net profit.

### 3.3 The NLEB model

A linear model was initially used to express the objective and the constraints, according to the farmers' perspective, which is the maximum profit. The model finds the optimal area for each crop (decision variables) per sub-watershed, such that the emission reduction targets for P or N are met, according to the regulations, the fertilizer application, while ensuring that the current levels of water use and available area for cultivation are not exceeded (Table 2).

Table 2. Mathematical description of the NLEB model (annual values): **1st Version**.

| Relation | Description |
|---|---|
| $x_c$, $c = 1,\ldots,28$. | Decision variables $X_c$ represent the area of each crop $c$ in [ha]. |
| $\max NP = \sum_{d \in D} \sum_{c \in C} (pr_c\, y_c - prod_{cost_c})\, x_{d,c}$ (11) | Objective function of Net Profit's (NP) maximization [CAD], as a function of each crop's product prices ($pr_c$ in [CAD/kg]), average yield ($y_c$ in [kg/ha]) and their typical production costs (prod_cost$_c$ in [CAD/ha]). $x_{d,c}$ expresses the crops' areas in [ha] as they are allocated per sub-watershed (d) |
| $\sum_{c \in C} x_{d,c} \leq TA_d$, $\forall d$ (12) | 1st constraint: not to surpass the available cultivated area per sub-watershed (Total Area: TA) [ha] |
| $\sum_{d \in D} \sum_{c \in C} wr_c\, x_{d,c} \leq TWU$ (13) | 2nd constraint: the typical water requirements for each crop ($wr_c$ in [m³/ha]) not to exceed the total water amount currently used for irrigation (TWU) [m³]. This constraint was applied for the whole range of TWU values from 0-50%, (each value representing a different Scenario), to explore the irrigation water and profits trade-offs |
| $\sum_{d \in D} \sum_{c \in C} fert_{jc}\, x_{d,c} \leq TF_{d,c}$ (14) | 3rd constraint: not to surpass the already implemented fertilization (TF$_c$) [kg] from each crop's requirement in fertilizers (fert$_c$) [kg/ha]. Three constraints were applied in this context, one for each of the following fertilizers, j = N, P$_2$O$_5$, K$_2$O |
| $\sum_{d \in D} \sum_{c \in C} P_c\, x_{d,c} \leq P_{des}$ (15) | 4th constraint: reducing P exports to desirable levels (P$_{des}$) [kg] from each crop's P export coefficients (P$_c$) [kg/ha]. This constraint was applied by |
| | using Scenarios for the whole range of reduction percentages of $P_{des}$ (from 0-50%, according to the Governmental goals) |
| $\sum_{d \in D} \sum_{c \in C} N_c \, x_{d,c} \leq N_{des}$ (16) | 5$^{th}$ constraint: reducing N exports to desirable levels ($N_{des}$) [kg] from each crop's N export coefficients ($N_c$) [kg/ha]. This constraint was applied by using Scenarios for the whole range of reduction percentages of $N_{des}$ (from 0-50%, according to the Governmental goals) |
| $y_c = \sum_{d \in D} y_c \, x_{d,c}$ (17) <br> $y_c^{min} \leq y_c \leq y_c^{max}, \; \forall c$ | The areas of the crops cannot change totally randomly, but in line with each sub-watersheds' production goals. Minimum and maximum y values were imposed to limit the impact of potential supply shocks to the market. A range of 50-150% of the current production was used, based on each crop's historically observed areas $x_{d,c}$ |

The examined policy suggests an optimization of the cropping distribution of the province's sub-watersheds, as mentioned in the study area section. Thus, the algorithm solves the problem 274 times. The analysis was carried out for the whole range of scenarios considering the reductions of TWU, $P_{des}$, $N_{des}$ (and their combinations) in order to provide the trade-offs among these parameters (representing the different policy goals) and the NP.

This model was also tested considering that the optimized crops' areas will translate to a different supply to the market, affecting thus the product prices. These changes driven by the optimized crop distributions were studied in a second version of this model. This modified version becomes non-linear, using the elasticity of the supply to the product prices, answering to the problem of reaching the Pareto frontier of the optimal trade-off between production and the associated market's behaviour. The modified version can be applied for both LKW and NLEB cases, and is described in the Appendix, as a computational addition of the main body of research outlined here.

### 3.4 The ESCA model

A Goal Programming (GP) model was developed to depict the situation of ESCA in a representative way for Irish agriculture and its animal farming concerns in general. The objectives set are: maximum sales, minimum production or capital costs, maximum exploitation of available area, minimum emissions of Phosphorus and Carbon, maximum organic fertilizer and minimum use of chemical fertilizer, maximum expected production, and minimum water use. The model allows for deviations (d) from one or more expected (or desirable) targets, and the policymakers can penalize those deviations as exceedances (+) or deficits (-) by weighting them (w: degree of undesirability of deviations). The model minimizes these deviations from the desirable goals, and also provides the optimal values of different animal types (decision variables), as well as for each goal set (Table 3). Three decision variables were considered, accounting for the main farming activities in Ireland: beef, dairy, poultry. The above (variables, goals and constraints) are indicative, as the user can easily omit goals and decision variables, or add more (e.g. crop types, or more animal types, minimize labour hours, maximum use of machinery, minimum transports, minimum N, etc.).

Table 3. Mathematical description of the ESCA model (annual values).

| Relation | Description |
|---|---|
| $X = beef, dairy, poultry$ | The decision variables: beef cows (beef: with index 1 in [heads]), dairy cows (dairy: with index 2 in [heads]) and poultry hens (poultry: with index 3 in [heads]) |
| Min Z = $\sum_{i=1}^{m} \sum_{k=1}^{K}(w_{i,k}^{+} d_i^{+} + w_{i,k}^{-} d_i^{-})$ (17) | The objective function minimizing the deviations (d) of the desirable Goals (i), weighted depending on their importance (w) |
| $s_1 \cdot beef + d_{s1}^{-} \geq TypicalSale_1$<br>$s_2 \cdot dairy + d_{s2}^{-} \geq TypicalSale_2$<br>$s_3 \cdot poultry + d_{s3}^{-} \geq TypicalSale_3$<br>(18) | Goal 1: Maximize the sales [€/year]. $s_1$, $s_2$, $s_3$ are the average earnings from livestock [€/head]. $TypicalSale_{1,2,3}$ are their respective Typical Sales [€] |
| $c_1 \cdot beef + c_2 \cdot dairy + c_3 \cdot poultry - d_c^{+} \leq$ Budget<br>(19) | Goal 2: Minimize the total production or capital cost. $c_1$, $c_2$, $c_3$ are the production or capital costs for livestock [€/head]. 'Budget' stands for an indicative expense scheduled to cover any production and capital costs [€] |
| $0.5 \cdot beef + 0.5 \cdot dairy + 0.01 \cdot poultry \leq$ AvailableArea<br>(20) | Goal 3: Do not exceed the available area (AvailableArea) [ha] (assuming 2 cows/ha gives 0.5 [ha/head] and 100 hens/ha gives 0.01 [ha/head]) |
| $e_1 \cdot beef + e_2 \cdot dairy + e_3 \cdot poultry - d_e^{+} \leq$ maxEmissionP<br>(21) | Goal 4: Minimize emissions of P, set to a maximum desirable value (maxEmissionP in [kg]). $e_1$, $e_2$, $e_3$ are the average P emissions for each decision variable [kg/head] |
| $ghg_1 \cdot beef + ghg_2 \cdot dairy + ghg_3 \cdot poultry - d_{ghg}^{+} \leq$ maxEmissionC (22) | Goal 5: Minimize emissions of C, set to a maximum desirable value (maxEmissionC in [kg]). This constraint is used to address the GHGs emissions issue. $ghg_1$, $ghg_2$, $ghg_3$ are the average C emissions for each decision variable [kg/head] |
| $of_1 \cdot beef + of_2 \cdot dairy + d_{of}^{-} - d_{of}^{+} <$ or $=$ or $> T_{of}$<br>(23) | Goal 6: Maximize the Organic Fertilizer for crops ($of_1$, $of_2$) in [kg/ha], compared to the produced value ($T_{of}$ in [kg]). This constraint is used to ensure the more efficient (and circular) fertilizer application, using first as much organic possible, and then adding chemical, if needed |
| $Cf_{required} \leq MaxChemical$<br>(24) | Goal 7: Minimize the chemical Fertilizer, under a desirable level (MaxChemical in [kg]). $Cf_{required}$ are the average fertilizer requirements for cultivation [kg/ha] |
| $water_1 \cdot beef + water_2 \cdot dairy + water_3 \cdot poultry - d_{water}^{+} \leq TWA$ (25) | Goal 8: Do not exceed the total water availability (TWA in [m³]) for animal use ($water_{1,2,3}$ in [m³/head]) |
| $y_1 \cdot rate \cdot beef + d_{y1}^{-} - d_{y1}^{+} <$ or $=$ or $>$ beefTarget<br>$y_2 \cdot rate \cdot dairy + d_{y2}^{-} - d_{y2}^{+} <$ or $=$ or $>$ dairyTarget | Goal 9: Reach the production targets (beefTarget, dairyTarget, cowTarget in [kg]) compared to their typical expected yields ($y_1$, $y_2$, $y_3$ in [kg/head]). The 'rate' expresses the growth of livestock to achieve their yield. e.g.: Assuming animals' weight and years to grow:<br>yield·rate· cow = [100kg/head]· [1/5y]· [head] = 20 kg/y |

$$y_3 \cdot \text{rate} \cdot \text{poultry} + d_{y3}^- - d_{y3}^+ < \text{or} = \text{or} > \text{poultryTarget} \quad \text{yield} \cdot \text{rate} \cdot \text{poultry} = [2.5 \text{kg/head}] \cdot [1/1y] \cdot [\text{head}] = 2.5 \text{ kg/y}$$

(26)

The model will provide the optimum herd size for poultry, beef and dairy cows, in order to achieve the minimum deviation from the environmental and economic targets set (quantifying them). More specifically, this approach can ensure the minimum deviation for expected sales, costs, organic fertilizer use, while the available water and area, chemical fertilizer, P and C emissions will not be exceeded. The data used were retrieved from official statistical databases, representative for the study area and Irish agriculture (see Table A3 of the Appendix). The weights (w) for each Goal's deviation were set by the analysts, as this is an indicative (demo) example, however, it is suggested defining them in a workshop with the local stakeholders having an interest to the problem's factors. Thus, it will be easier for them to understand the problem, its assumptions and trade-offs, reaching easier to an agreed plan (social acceptance) for implementation.

## 4. Results

The methodologies described in the previous sections were developed and solved in Python (scripts provided in Appendix). This work aimed to demonstrate these optimization set-ups as approaches for working with similar problems, so certain parameters defined by the analysts, can ideally be estimated through modelling (e.g. economic, hydrological, agronomic models).

### 4.1 LKW model

Since the major concern of this region is the availability of water for irrigation, the scenarios focus on decreasing water use without compromising profits for farmers. Initially, the model is solved for a zero water use reduction to determine what would be the most profitable way to allocate the current water resources in irrigation. Then a 20 and 40% reduction in water use is implemented.

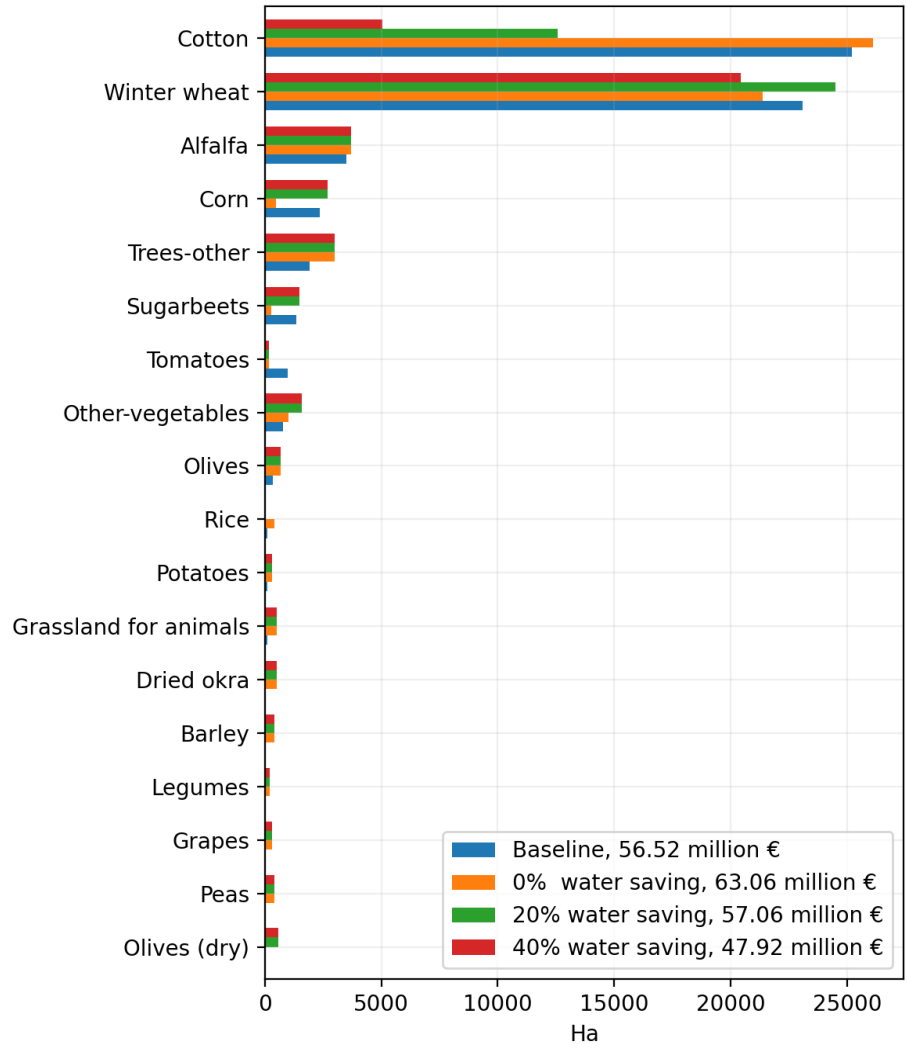

Figure 2. Crop areas for the baseline and the optimal solutions for water use reduction of 0, 20, 40%. The label shows the total annual profit for the crop selection.

The optimal results indicate that there are less water consuming crop distributions that can maintain the current NPs for farmers. The optimal solution for the same water use as the baseline's shows that a more profitable crop selection can be achieved, increasing the profits by 11%. For a 20% water use reduction, an alternative crop distribution can again achieve slightly higher profits for farmers. Finally, the scenario of reducing by 40% the water use show that it is still possible to avoid sacrificing a big proportion of the profits, as in this case we have only decrease of 15%.

In every scenario, the other constraints are reaching their threshold values, namely the same amount of labour hours, fertilizer and water use, as in the baseline are maintained.

**4.2 NLEB model**

The main concern of the NLEB case is water quality, not quantity. P and N exports from agricultural runoff were examined under the baseline and a range of scenarios, as mentioned. Table 4 shows the baseline values of the crops' production, P and N exports. As a first step, the optimization model was solved allowing the same total maximum nutrient exports as in the baseline to show the effect that an alternative (optimal) crop distribution can bring to the production, and thus profits. The results show there is a gain of 36%, by only changing the crop distribution.

Table 4. Crop production, P and N exports for the baseline and the optimum solution, allowing the same total maximum export of P and N as in the baseline.

| Crops | Production [tons/year] | | P exports [tons/year] | | N exports [tons/year] | |
|---|---|---|---|---|---|---|
| | Baseline | Optimal | Baseline | Optimal | Baseline | Optimal |
| Total corn | 9,981,214 | 6,674,986 | 3,953 | 2,644 | 8,430 | 5,637 |
| Alfalfa and alfalfa mixtures | 3,665,686 | 5,498,530 | 1,112 | 1,668 | 2,973 | 4,459 |
| Soybeans | 3,549,096 | 3,996,848 | 10,547 | 11,878 | 8,864 | 9,982 |
| Total dairy | 2,364,101 | 1,403,045 | 1,612 | 957 | 1,269 | 753 |
| All other tame hay and fodder crops | 2,056,044 | 3,084,067 | 295 | 442 | 1,576 | 2,364 |
| Tomatoes | 530,059 | 795,088 | 57 | 85 | 48 | 72 |
| Potatoes | 408,591 | 612,886 | 39 | 58 | 66 | 99 |
| Sugar beets | 238,364 | 357,547 | 8 | 12 | 21 | 31 |
| Oats | 155,221 | 77,610 | 206 | 103 | 134 | 67 |
| Barley | 154,139 | 231,208 | 85 | 128 | - | - |
| Apples total area | 141,173 | 211,760 | 36 | 53 | 25 | 38 |
| Mixed grains | 111,150 | 55,575 | 183 | 91 | 312 | 156 |
| Total rye | 58,667 | 29,334 | 43 | 22 | 44 | 22 |
| Other field crops | 47,617 | 71,425 | 32 | 49 | 46 | 68 |
| Dry white beans | 44,019 | 66,029 | 21 | 31 | 70 | 105 |
| Pumpkins | 43,617 | 65,426 | 3 | 4 | 6 | 10 |
| Cucumbers | 36,015 | 54,023 | 5 | 8 | 5 | 7 |
| Canola (rapeseed) | 25,912 | 38,868 | 7 | 10 | 25 | 38 |
| Cabbage | 23,055 | 34,583 | 2 | 4 | 2 | 3 |
| Ginseng | 9,729 | 14,594 | 9 | 14 | 16 | 24 |
| Greenhouse vegetables | 7,096 | 10,645 | 2 | 2 | 2 | 2 |
| Forage seed for seed | 5,273 | 7,909 | <1 | <1 | <1 | 1 |
| Dry field peas | 3,407 | 1,704 | 12 | 6 | 12 | 6 |
| Other greenhouse products | 1,242 | 1,862 | <1 | <1 | <1 | <1 |
| Flaxseed | 117 | 176 | <1 | <1 | <1 | <1 |
| Mustard seed | 116 | 174 | <1 | <1 | <1 | <1 |
| Sunflowers | 65 | 32 | <1 | <1 | <1 | <1 |

This finding is further demonstrated in Figure 3, for all scenarios explored, namely the whole set of values' combinations for P and N exports reduction, both ranging from 0-50%. Figure 3 shows how the profits are changing (Δutility) compared to the baseline value (zero-point), under all different levels of P and N reduction percentages. Taking into account all these combinations the model produced a total of 676 runs. The results show that P exports can be reduced up to 42% and N exports can be reduced up to 46% while the profits from agriculture can remain stable or become higher, just be alternating the crop distribution.

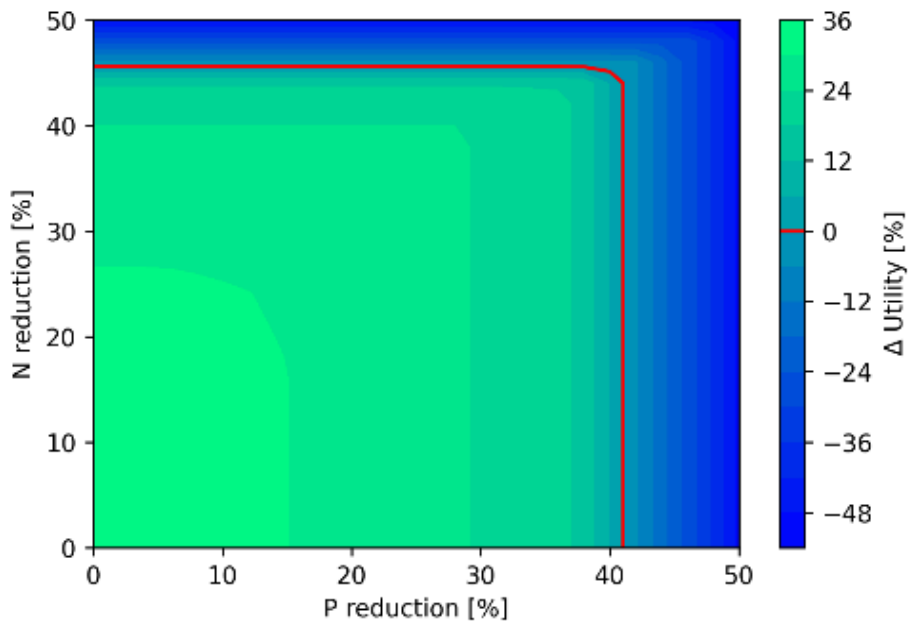

Figure 3: Change of profits from agriculture (Δutility) from the baseline for varying P or N reductions using the model's version 1. The red line shows the contour where the utility is the same as in the baseline.

The constraints were again maintained at the same levels as in the baseline situation, while the effect of the optimized results of each scenario can be further explored for each crop. Commenting on the agricultural production of the region with reduced P and N exports, it was found that the same or higher production can be achieved. Indicatively with respect to the main crops, corn's production could be increased up to 20-30% P and N exports reduction, for dairy 20-40%, alfalfa, tomatoes and potatoes even more than 40%, and soybeans up to 5-10%. It is worthy to mention that these values in reality can be further improved by applying BMPs, or other practices supporting farming and efficiency and pollution minimization.

### 4.3 The ESCA model

Three scenarios were created following different preferences on the importance of achieving the economical or environmental targets of the model. These preferences are reflected by the penalization given to the deviations from the target values, i.e. each scenario differs only in the $w_i$'s assigned to each goal (Table 5). The scenarios created are the following:

a) Scen. A: Extremely environmentalist (only caring to have zero emissions and ignoring the other goals).
b) Scen. B: Intensive farmer (only caring about maximum sales and profits, and then reduced costs while ignoring all the rest).
c) Scen. C: A balanced penalization, representing the 'middle solution' between the first two scenarios.

Table 5. Weights assigned per scenario (custom scale 0-1).

| Deviations penalized | Scen.A | Scen.B | Scen.C |
|---|---|---|---|
| Deficit of beef sales ($d_{s1}^-$) | 0.0 | 1.0 | 0.5 |
| Deficit of dairy sales ($d_{s2}^-$) | 0.0 | 1.0 | 0.5 |
| Deficit of poultry sales ($d_{s3}^-$) | 0.0 | 1.0 | 0.5 |
| Exceedance of costs – budget ($d_c^+$) | 0.0 | 0.1 | 0.0 |
| Exceedance of P emissions ($d_e^+$) | 1.0 | 0.0 | 0.5 |
| Exceedance of C emissions ($d_{ghg}^+$) | 1.0 | 0.0 | 0.5 |
| Exceedance of organic fertiliser ($d_{of}^+$) | 1.0 | 0.0 | 0.5 |
| Deficit of organic fertiliser ($d_{of}^-$) | 0.0 | 0.0 | 0.0 |
| Exceedance of beef production ($d_{y1}^+$) | 0.01 | 0.0 | 0.005 |
| Deficit of beef production ($d_{y1}^-$) | 0.01 | 0.0 | 0.005 |
| Exceedance of dairy production ($d_{y2}^+$) | 0.01 | 0.0 | 0.005 |
| Deficit of dairy production ($d_{y2}^-$) | 0.01 | 0.0 | 0.005 |
| Exceedance of poultry production ($d_{y3}^+$) | 0.01 | 0.0 | 0.005 |
| Deficit of poultry production ($d_{y3}^-$) | 0.01 | 0.0 | 0.005 |
| Water deficits ($d_{water}^+$) | 0.05 | 0.0 | 0.05 |

A cultivated area of 4000 ha was assumed, with a Budget constraint of 150,000 €/year, and 0.2hm³ annual water availability. The typical P and C emission targets are defined by European or national (local) policies as thresholds. In this example, these were set to 3,350 kg/year and 45,000 kg/year respectively.

Scenario A (Scen.A) sets the criterion of controlling emission and pollution, so it prefers poultry farms, followed by beef and dairy cows. There are some losses in the sales of poultry, beef and dairy products, compared to the expected targets (e.g. current averages). The P and C emissions above the target (exceedances) are zero, as well as the water deficit and the organic fertiliser.

Scenario B (Scen.B) on the contrary, having primarily economic motives, prefers beef and dairy cows, and choses to surpass the budget constraint (by setting lower $w_i$) in order to over-produce and exceed supply and sales, aiming to higher profits. However, this results in significant water deficits and exceedance of P emissions.

Scenario C (Scen.C) as a balanced approach considers all three animal types: beef, poultry and then dairy cows. This achieves all the environmental objectives of Scen.A (P, C, water, organic fertiliser), with less losses of sales for beef compared to Scen.A, and also less losses of poultry sales compared to Scen.B. All the production targets are met, as well as the budget constraint, unlike Scen.B, so economic objectives can be also satisfied.

Table 6. Results of the ESCA model per scenario.

| Parameters | Scen.A | Scen.B | Scen.C |
|---|---|---|---|
| Beef (Heads) | 71 | 200 | 130 |
| Dairy (Heads) | 48 | 181 | 48 |
| Poultry (Heads) | 107 | - | 107 |
| Loss in beef sales (€/year) | 16,099 | - | 8,798 |
| Loss in dairy sales (€/year) | 19,941 | - | 19,941 |
| Loss in poultry sales (€/year) | 13,932 | 15,000 | 13,932 |
| Exceedance of costs (€/year) | - | 205,893 | - |
| Exceedance in emissions of P (kg/year) | - | 1,002 | - |
| Exceedance in emissions of C (kg/year) | - | - | - |
| Exceedance of Organic Fertilizer (kg/year) | - | 8,758 | - |
| Deficit of Organic Fertilizer (kg/year) | 7,224 | - | 3,895 |
| Exceedance in beef supply (kg/year) | - | 1,128,200 | 511,612 |
| Deficit in beef supply (kg/year) | - | - | - |
| Exceedance in dairy supply (kg/year) | - | 1,552,734 | - |
| Deficit in dairy supply (kg/year) | - | - | - |
| Exceedance in poultry supply (kg/year) | - | - | - |
| Deficit in poultry supply (kg/year) | - | 156,000 | - |
| Water deficits (m$^3$/year) | - | 2,358,911 | - |

Analyzing different factors in the same model is challenging, however a more thorough picture of trade-offs can be seen under different scenarios representing different stakeholders. The results of Table 6 reflect the importance of planning multiple (conflicting) objectives together as a system in order to provide sustainable solutions. Such solutions are feasible, and the learnings from such models can support policymakers and practitioners to better understand complex systems. Similar approaches are encouraged in terms of building integrated databases that will lead to a holistic monitoring-modelling of the system as a whole, ensuring that no discipline will act at the expense of another.

### 5. Limitations and Future research

Unavoidably, any model that attempts to describe real situations cannot be perfect. Starting from the purpose of the examples presented, they aimed to demonstrate how simple optimization set-ups could be used to address various problems and achieve optimum solutions. The aim was not to propose a specific agricultural plan that must be followed by farmers in each case study, rather than tools to take into consideration as many factors as possible and combines them to find more sustainable and profitable solutions. An inherent limitation in such exercises is the data availability and quality. Trying to depict the situation in three rural areas, with the two of them hardly being monitored, and of small scale, is ambitious. Precise input data in agriculture are rarely if ever available, and this can affect the models' outputs. NLEB had the more complete and organized data, and this allowed us to develop two versions of this model, taking into account several crops and parameters. However, in all three cases most data used are obtained from official databases, all of them were validated with locals, and are based on average annual values, so to have representative and accurate models, as possible. In the future, specific parameters can be defined considering the case- and market- based features of each study area, if the data capacity and transparency increases. Thus, more flexible tools can be developed, able to better cope with the temporary nature of most data used, that are constantly changing.

Another limitation arising from lack of data, is that no spatial optimization for the proposed crop distributions was performed. In a later stage, however, where data can be found, this would be highly valuable and included in our plans. Moreover, in this study, no local stakeholder analysis was carried out to weight the goals and evaluate their importance, because the aim at this point was to describe the methods. However, this is absolutely required in future studies, and we have already started such efforts (initially for LKW) using the learnings and knowledge obtained from this work. Stakeholder are keen on learning how multiple conflicting objectives can be modelled together and provide useful information. Also, the proposed models are flexible, as users can modify the problems, add variables, constraints, etc. Through such engagement processes, additional measures for the more efficient resource management (BMPs) can be better targeted and implemented to further improve the systems' functions.

A final limitation that we identified is that instead of using typical average values for certain factors, ideally, we would like to base them on actual hydrologic, agronomic, bio-economic models (e.g. water availability and use, fertilizer and nutrient exports, farmers and livestock economics, etc.). A two-way information between hydrologic, agronomic, bio-economic models and the optimization planning models presented should be followed to complete the design-implementation frame, and majorly expand the capabilities in planning. More specifically, time-series analyses, forecasts, examination of more physical scenarios (e.g. extremes, climate change, economic externalities, etc.) can be tested. Unfortunately, this could not be presented in the length of this research paper, which already includes a lot of compressed information, however we are planning to analyze one case-study more thoroughly in the near future.

The expansion and coupling of the models presented in this study with environmental models in particular (e.g. Soil and Water Assessment Tool – SWAT, or GIS-based models, that use Python) is included in our future plans, and for that reason our three examples were developed in Python, in order to be easily compatible.

Of course, models provide useful answers, but not all the answers. The mindset of approaching several problems with conflicting objectives, in an integrated and inter-disciplinary way is the element highlighted most in this work. A message is that even simple models and analyses can provide significant insights and if applied right, can be highly informative, so any application is encouraged, especially in poorly managed sites.

## 6. Concluding remarks

In this study, three complex problems of agricultural water management were examined, in an attempt to cover the most common concerns around the topic. Dry and crop-intensive areas, such as LKW, find it increasingly difficult to cover the irrigation water demand. Rural areas that do not face water scarcity issues, usually focus on water quality protection and improvement. The main pressures can be fertilizers or animals, like in NLEB and ESCA, respectively. In most cases, environmental conservation and improvement competes with agriculture's economic goals. The achievement of productive goals under environmental constraints reminds the definition of the economic problem, which in general tries to cover increasing needs with limited resources. Although there is no "one size fits all" approach, a common element is used in all models presented, i.e. the maximization of profits from agriculture, while water use and pollution are set to desirable thresholds, estimating of their trade-offs. This approach (either as linear, non-linear, or multi-objective optimization) provided encouraging results for every case, in the form of alternative cropping plans or combinations of desirable goals. The results show that agricultural economies can be prosperous under environmental constraints, which can be tested under different conditions (scenarios) allowing thus to the decision-makers to explore the trade-off of 'conflicting' parameters, understand how the system works, and finally agree on a plan. This social acceptance can be achieved by examining the different scenarios, as shown, and through the weights $w_i$ for the ESCA case. Such processes can provide a basis for future collaborative planning between government and stakeholders, towards economically effective and environmentally sustainable management.

The parameters examined were chosen to depict the actual concerns of most policies, for example: the European Water Framework Directive, with its River Basin Management Plans, the Common Agricultural Policy, Nitrate Policies, Canada's Action Plans, Sustainable Development Goals, etc. All those plans have common goals in a degree, support sustainability, integrated management and multi-disciplinary approaches to balance similar issues. This requires integrated databases and systemic understanding, which can be achieved through similar modelling processes. Usually, policy-makers are not likely to accept a new modelling approach, or even modelling itself, unless it is obvious that it will improve the performance of their work and help them address problems they are trying to solve (Loucks and van Beek, 2017). From our experience in all three case-studies, stakeholders tend to seek more scientific approaches, to balance conflicts and achieve multiple benefits. The transition to a multidisciplinary world with the modernization of traditional management practices requires respective knowledge and capacity, and this is increasingly highlighted by international (and national) policy agendas. These elements are optimistic for bringing knowledge out of the academic environment and collaborate towards a more sustainable future.